\documentclass[draft]{agujournal2019}
\usepackage[normalem]{ulem}
\usepackage{url} 
\usepackage{lineno}
\usepackage[inline]{trackchanges} 
\usepackage{soul}
\usepackage[numbers,square]{natbib}
\usepackage{multirow}
\usepackage{tabularx}
\usepackage{booktabs}
\usepackage{xcolor}
\draftfalse

\journalname{QJRMS}

\begin{document}

%
%



\title{Assimilation of machine learning-predicted nitrate to improve the quality of phytoplankton forecasting in the shelf sea environment.}

%
%




\authors{Deep S. Banerjee$^{1,2}$, Jozef Sk\'akala$^{1,2}$, David Ford$^{3}$}


\affiliation{1}{Plymouth Marine Laboratory, Plymouth, UK}
\affiliation{2}{National Centre for Earth Observation, Plymouth, UK}
\affiliation{3}{Met Office, Exeter, UK}





\correspondingauthor{Jozef Sk\'akala}{jos@pml.ac.uk}


\vspace{16.6cm}



%
%

%
%


\begin{abstract}
We demonstrate that assimilating Neural Network (NN)-predicted surface nitrate leads to a major improvement in phytoplankton short-range (1-5 day) dynamical model forecasts for the North-West European Shelf (NWES) seas. We show that assimilation of only ocean color chlorophyll-$a$ in the current Met Office NWES operational system can lead to excess surface nitrate concentrations in the post-Spring bloom period and these are a major reason behind some known, fast-growing biases in NWES phytoplankton forecasts during late Spring and Summer. Assimilating observations of nitrate would potentially help address this, but NWES nitrate data are typically not available in sufficient abundance to be effectively assimilated. 
We have therefore used a recently developed and validated neural network (NN) model predicting surface nitrate concentrations from a range of observable variables and assimilated the NN-predicted nitrate within a research and development version of the Met Office's NWES operational forecasting system. As a result of nitrate assimilation the phytoplankton 5-day forecast skill improves by up to 30\%. We show that although much of this improvement can be achieved by using a weekly nitrate climatology predicted by the NN model, there is a clear advantage in using flow-dependent nitrate data. We discuss the impacts of this improvement on a range of additional eutrophication indicators, such as dissolved inorganic phosphorus and sea bottom oxygen. We argue that it should be feasible to upgrade this approach to a fully hybrid machine learning - data assimilation within the near-real time NWES operational forecasting system.
\end{abstract}


%
%

%


%
%
%
%

\section{Introduction} 

Operational monitoring and forecasting of marine biogeochemistry can provide an essential source of information for water quality, fisheries and aquaculture management, as well as climate mitigation and adaptation planning and policy (e.g., \cite{fennel2019advancing}). One of the regional seas monitored and predicted by operational systems is the North West European Shelf (NWES). NWES plays a significant role in global biogeochemical cycles, i.e. it acts as a sink for atmospheric CO$_{2}$ through biological productivity and eventually exports organic matter to the open ocean, influencing the global carbon budget \citep{legge2020carbon}. Additionally, efficient nutrient cycling through riverine discharge, atmospheric deposition and shallow bathymetry supports the maintenance of a diverse marine ecosystem, making NWES vital for the European economy \citep{pauly2002towards}.

NWES ecosystem is characterized by strongly seasonal dynamics: i.e. during spring, the sunlight, onset of stratification, and abundance of nutrients, accumulated over the last winter near the ocean surface through mixing by storms, set up ideal conditions for phytoplankton to bloom. This spring bloom contributes a major portion of annual primary production (e.g., \cite{silva2021twenty, gonzalez2022onset}). During the bloom, phytoplankton, especially diatoms, start rapidly assimilating nutrients, causing a sharp decrease in nutrient concentrations in the upper oceanic layer. Through this process, the surface water becomes eventually depleted in nutrients (especially nitrate and phosphate), which limits phytoplankton growth. This continues until Autumn, since strong stratification of the water column separates the surface from the nutrient-rich water below the pycnocline, with phytoplankton growth often limited to deep chlorophyll maxima occurring around/under the pycnocline (e.g., \cite{weston2005primary, skakala2021towards, loveday2021daily}). However, sometimes in late summer or early autumn, due to wind-driven mixing (often due to storms), the stratification breaks down and nutrients upwell from deeper water to the surface causing a secondary bloom \citep{capuzzo2018decline}. Then, during winter, strong persistent winds cause stratification to break down completely, cooling the ocean surface and increasing the water density, which triggers convective mixing \citep{sharples2006interannual}. As a result of strong winter mixing, the remineralised nutrients accumulated during late summer and autumn get upwelled near to the surface setting up conditions for the next spring bloom \citep{lohse1995sediment}. 

The Met Office runs an operational physics-biogeochemistry forecasting system for the NWES, each day producing forecasts with up to 6-day lead-time for a range of key variables, such as phytoplankton biomass, nutrients, oxygen and underwater visibility (see https://\-www.metoffice.gov.uk/\-services/\-data/\-met-office-marine-data-service, the Copernicus Marine Service and also \cite{skakala2025marine}). These forecasts provide an early indicator for the risk of eutrophication, a recurring problem in parts of NWES (e.g., \cite{axe2017eutrophication, devlin2023first}), with impact on aquaculture operations and coastal management. Furthermore, the underwater visibility forecasts inform underwater operations (e.g., \cite{skakala2025marine}), with other applications for the biogeochemistry forecasts including navigating autonomous observing platforms \citep{ford2022solution}, and potentially providing useful information to models predicting toxic algal blooms. However, a key operationally forecasted variable, phytoplankton biomass, tends to be overestimated in the forecast, with the positive bias growing with forecast lead time \citep{skakala2018assimilation}.
These biases are largest in the late Spring-Summer and can impact the forecast quality of other biogeochemical variables, due to the central role of phytoplankton in the marine ecosystem.


The phytoplankton forecast biases can be understood based on the interaction between the model dynamics and the data assimilation design. The NWES biogeochemistry model, the European Regional Seas Ecosystem Model (ERSEM, \cite{baretta1995european, butenschon2016ersem}) has major seasonal phytoplankton chlorophyll-$a$ biases relative to both satellite and in situ data. It tends to substantially overestimate phytoplankton concentrations during the bloom period, with the bloom onset often modelled too late (e.g., \cite{kay2019north, skakala2018assimilation, skakala2020improved, skakala2022impact}). The exact causes of these seasonal biases in ERSEM phytoplankton are not entirely known, and are likely due to a complex interaction of multiple drivers, including zooplankton grazing (e.g. see \citep{skakala2020improved, skakala2021towards} for some discussion). The biases in chlorophyll-$a$ concentrations are corrected by assimilating satellite ocean color-derived surface chlorophyll into the model, substantially lowering phytoplankton concentrations during the bloom in the analysis, relative to the free run. However, due to a lack of other available observations, ocean colour-derived chlorophyll is the only biogeochemical variable currently assimilated operationally. The data assimilation (DA) scheme used only directly updates the phytoplankton size-class chlorophyll and biomass variables \citep{skakala2018assimilation}, meaning that other biogeochemical variables are not directly constrained by observations.
Due to the routine reduction of phytoplankton biomass by DA during the Spring bloom, productivity and therefore nutrient uptake is lower in the analysis, meaning surface nutrients do not get depleted, as they would get in the free run \citep{banerjee2025improved, kay2019north}. This means, in late Spring-Summer, just after the end of the bloom, when the nutrients in the model free run are exhausted and phytoplankton growth becomes nutrient-limited, the analysis continues to see availability of nutrients combined with the good light-conditions characteristic for this season.  During the forecast, when the model is no longer constrained by DA, these conditions trigger rapid phytoplankton growth, beyond the period in which this would be otherwise present in the free run. This results in positive NWES-wide biases in the forecast phytoplankton concentration developing during this period, degrading the forecast skill with lead time. These widespread biases mean the reduction in forecast skill with lead time is much greater than would be expected simply through the system's chaotic dynamics.

In recent decades, machine learning (ML) has emerged as a transformative tool to address several challenges across a wide range of fields, from healthcare and finance to environmental sciences. ML algorithms are a branch of artificial intelligence that learn patterns and trends from datasets and resolve non-linear and complex relationships that are not immediately apparent. ML has already become a vital part of marine science (e.g. \cite{sonnewald2021bridging}), including having impacts on operational oceanography (e.g., \cite{kochkov2021machine, heimbach2024crafting}) and marine biogeochemistry modelling (e.g., \cite{mattern2013sensitivity, schartau2017reviews, skakala2023future, wu2025neural}). A neural network (NN) model has been recently developed by \cite{banerjee2025improved} to predict gap-free surface nitrate concentrations on the NWES from a set of routinely observed variables. This model, trained on in situ nitrate observations from the International Council for the Exploration of the Sea (ICES) database (https://www.ices.dk), was demonstrated to be highly skilled in reproducing independent in situ data, albeit with slightly coarsened spatial and temporal effective resolution \citep{banerjee2025improved} (for more detail see Sec.2.2). The work of \cite{banerjee2025improved} presents us with a new opportunity to tackle the phytoplankton forecasting problem, i.e. if nitrate concentrations predicted from the observable variables were assimilated into the model together with the (already assimilated) satellite chlorophyll, they could effectively correct nitrate alongside phytoplankton biomass. Since nitrate is a key limiting nutrient on the NWES (e.g. \cite{axe2017eutrophication, devlin2023first}), we anticipate (and consequently demonstrate) that correcting the nitrate biases by assimilating the NN-predicted nitrate will have a major positive impact on the phytoplankton forecast skill. The approach undertaken here could be presented as a form of ML-bias correction of a biogeochemistry model, and broadly understood within the area of combined ML-DA approaches, a subject that has become very popular in recent years (e.g. see the review by \cite{cheng2023machine}). It should be said that in situ nitrate concentrations (both real and synthetic observations) have been already assimilated in the past into marine biogeochemistry models \citep{anderson2000physical, ourmieres2009key, ford2021assimilating}, furthermore work deriving nitrate concentrations using ML is not entirely new \citep{sauzede2017estimates, chen2023estimation}, including assimilating ML-derived nitrate from Bgc-Argo floats in the Mediterranean Sea \cite{amadio2024combining}. However, unlike the previous studies, here we assimilate flow-dependent (evolving in real time) gridded gap-free surface nitrate data, allowing for rapid, domain-wide improvements to the model forecasting capability in the mixed layer. 


In this study, we conducted three experiments: (i) a reference run with a set-up reasonably close to the one used operationally, i.e. including assimilation of physics observations and ocean color-derived chlorophyll (as no nitrate was assimilated in this experiment, it will be further labeled as ``no-nit DA''), (ii) an experiment additionally assimilating flow-dependent ML-generated nitrate data (labeled ``ML-nit DA''), (iii) an equivalent experiment to experiment (ii), but assimilating a climatology of those ML-generated nitrate data instead of the flow-dependent nitrate  (labeled ``clim-nit DA''). 
The last experiment enables us to assess how much of the phytoplankton forecast improvement can be achieved by a form of relaxation towards the nitrate seasonal climatology values, and how important it is to have a flow-dependent ML nitrate prediction. All three experiments were performed for the biologically productive period (March-September) of 2018. 

In this study, the nitrate data were generated "offline", meaning the ML-generated nitrate data were separately predicted from a previously-run reanalysis rather than as part of the assimilation step.
However, it should be noted that the reanalysis used was produced using a broadly similar model and assimilation set-up to that used in this study, ensuring a reasonable level of consistency.
It should also be noted that the longer-term objective is to develop an "online" setup: In this future framework, at each assimilation time the system ideally creates in its first step the analysis using the data assimilated in the current version of the system (i.e. physics data and satellite chlorophyll-$a$) and then it will use a selected set of analysis variables (e.g. SST, chlorophyll-$a$) as inputs into the ML model to predict surface nitrate. In the second step this predicted nitrate will be assimilated into the model (correcting only the nitrate, which was unchanged during the first assimilation step), further updating the analysis state. If this scheme added too much computational expense relative to the benefit of predicting the nitrate from the analysis state (analysis state being the most reliable estimate of the ocean at the given day and consistent with how the ML model was trained), we could also run an alternative, where we use the inputs from the background state, run the ML model first and then perform only one assimilation step for both nitrate and the currently assimilated data. Another, computationally efficient possibility is to predict the nitrate directly from the available observations at each assimilation step. Either way, we argue that our experiments demonstrate the potential feasibility of running the system in this online mode.

\section{Methodology}
\subsection{Met Office operational biogeochemical forecasting system} 

The Met Office runs an operational forecasting system for NWES biogeochemistry \citep{edwards2012validation, kay2019north}, with products made freely available for a range of users (https://www.metoffice.gov.uk/\-services/data/met-office-marine-data-service and Copernicus Marine Service). This uses the hydrodynamic model Nucleus 
for European Modelling of the Ocean (NEMO, \cite{madec2015nemo}) coupled with ERSEM \citep{baretta1995european, 
butenschon2016ersem}, through the Framework for Aquatic Biogeochemical Models (FABM, \cite{bruggeman2014general, fabm}). The system assimilates data into the model using the variational DA software NEMOVAR \citep{mogensen2009nemovar, mogensen2012nemovar}.

\subsubsection{The physical model}

The physical model NEMO is a finite difference, hydrostatic, primitive equation ocean general circulation model \citep{madec2015nemo}. 
The NEMO configuration used in this study is very similar to e.g. \cite{skakala2021towards, skakala2022impact, skakala2024how} and has been described therein: it uses the CO6 NEMO version, based on NEMOv3.6, a development of the CO5 configuration explained in detail by \cite{odea2017co5}. The model has approximately 7 km spatial resolution on 
the Atlantic Margin Model (AMM7) domain using a terrain-following $z^{*}-\sigma$ coordinate system with 51 vertical levels \citep{siddorn2013analytical}. In these experiments the lateral boundary conditions for physical
variables at the Atlantic boundary were taken from the previously operational Met Office North Atlantic deep ocean model \citep{storkey2010forecasting} and the Baltic boundary conditions from the Copernicus Marine Service operational Baltic Sea model \citep{berg2012implementation}.
We have used river discharge based on data from \cite{lenhart2010predicting}. The atmospheric forcing came from the Met Office Unified Model global numerical weather prediction system \citep{tonani2019impact}. 

\subsubsection{The biogeochemistry model}

ERSEM is a lower trophic level ecosystem  model
based on  pelagic  plankton,  and  benthic fauna \citep{blackford1997analysis}. The model divides autotrophs into four phytoplankton functional
types (PFTs) largely based on their size \citep{baretta1995european}: picophytoplankton, nanophytoplankton, diatoms and dinoflagellates. ERSEM uses variable stoichiometry for the simulated plankton groups 
\citep{baretta1997microbial, geider1997dynamic} and each PFT biomass is represented in terms of chlorophyll, carbon, nitrogen and phosphorus, 
with diatoms also represented by silicon.  ERSEM predators are represented by three zooplankton types (mesozooplankton, microzooplankton and heterotrophic nanoflagellates), 
with organic material being decomposed by one functional type of heterotrophic bacteria \citep{butenschon2016ersem}. The ERSEM inorganic component consists of nutrients 
(nitrate, phosphate, silicate, ammonium and carbon) and dissolved oxygen. The carbonate system is also included in the model \citep{artioli2012carbonate}. 

\subsubsection{The data assimilation system}

NEMOVAR is used here in a 3DVar configuration \citep{mogensen2009nemovar, mogensen2012nemovar, waters2015implementing} and its specific implementation in the NWES system has been described in a range of recent papers, e.g. see \cite{king2018improving, tonani2019impact} for the physics, and \cite{ford2022solution, fowler2023validating, kay2019north, skakala2018assimilation, skakala2020improved, skakala2021towards, skakala2022impact, skakala2024how} for the biogeochemistry. NEMOVAR uses First Guess at Appropriate Time (FGAT), which is applied to calculate the innovations between the observed values and model background at the nearest model timestep to the observation times, during a 24 hour forecast. Then NEMOVAR is used to produce a set of increments to update the model state variables. The increments are added into the model gradually over the same 24 hours to avoid generating sudden shocks, using incremental analysis updates (IAU, \cite{bloom1996data, king2018improving, waters2015implementing}). In the physical DA application, NEMOVAR applies balancing relationships within the assimilation step and delivers a set of increments for temperature, salinity, sea surface height (SSH) and the horizontal velocity components. In its biogeochemical application it calculates a set of increments separately for each assimilated variable and in specific cases balancing relationships are subsequently used to distribute those increments into a selected range of other ecosystem model variables.


In the operational NWES forecasting context NEMOVAR assimilates with a daily cycle sea surface temperature (SST), satellite sea level anomaly, temperature and salinity profiles, and satellite ocean-color derived total (log) chlorophyll-$a$. Although not used in operational forecasting, other options are available in the system, i.e. PFT (log) chlorophyll-$a$ assimilation \citep{skakala2018assimilation} (used in Copernicus reanalysis and also here, see Sec.2.3), assimilation of PFT absorption \citep{skakala2020improved} and assimilation of data from gliders \citep{ford2021assimilating, skakala2021towards}. In unpublished experiments (see https://\-meetingorga\-nizer.copernicus.org/\-EGU25/\-EGU25-14292.html), assimilation of nitrate (measured and ML-derived), chlorophyll-$a$ and oxygen from BGC-Argo and ships has also been established.

In the (log) chlorophyll-$a$ assimilation NEMOVAR is used to calculate increments to surface chlorophyll-$a$. When PFT chlorophyll-$a$ is assimilated increments are directly calculated for each PFT; when total chlorophyll-$a$ is assimilated the increments to total chlorophyll-$a$ produced by NEMOVAR are converted to increments to PFT chlorophyll-$a$ based on the forecast (background) PFT-to-total chlorophyll ratio at each grid point.
The increments are further propagated to other PFT biomass components (carbon, nitrogen, phosphorus, silicon) based on forecast PFT stoichiometry. 

NEMOVAR uses for both physics and biogeochemistry externally supplied, spatio-temporally varying observation error variances (the observation error correlations are unaccounted for). The chlorophyll-$a$ background error variances are based on the work of \cite{skakala2018assimilation} using ensemble simulations from \cite{ciavatta2016decadal}. The physics background error variances are taken from \cite{king2018improving}. For both biogeochemistry and physics variables the same horizontal correlations are used, as described by \cite{king2018improving}.
For physics variables vertical correlations are as described by \cite{king2018improving}, based on flow-dependent vertical length scales, which are a linear function of depth until the base of the mixed layer and then scale with the spacing of the vertical layers in the model grid (for details see Eq.1 in \cite{skakala2021towards}). For biogeochemistry, in this study, NEMOVAR was just used to calculate surface increments, which were then applied equally throughout the model mixed layer.

\subsection{Nitrate data assimilation based on a neural-network prediction}

\begin{figure}
\hspace{2cm}
\noindent\includegraphics[width=9cm]{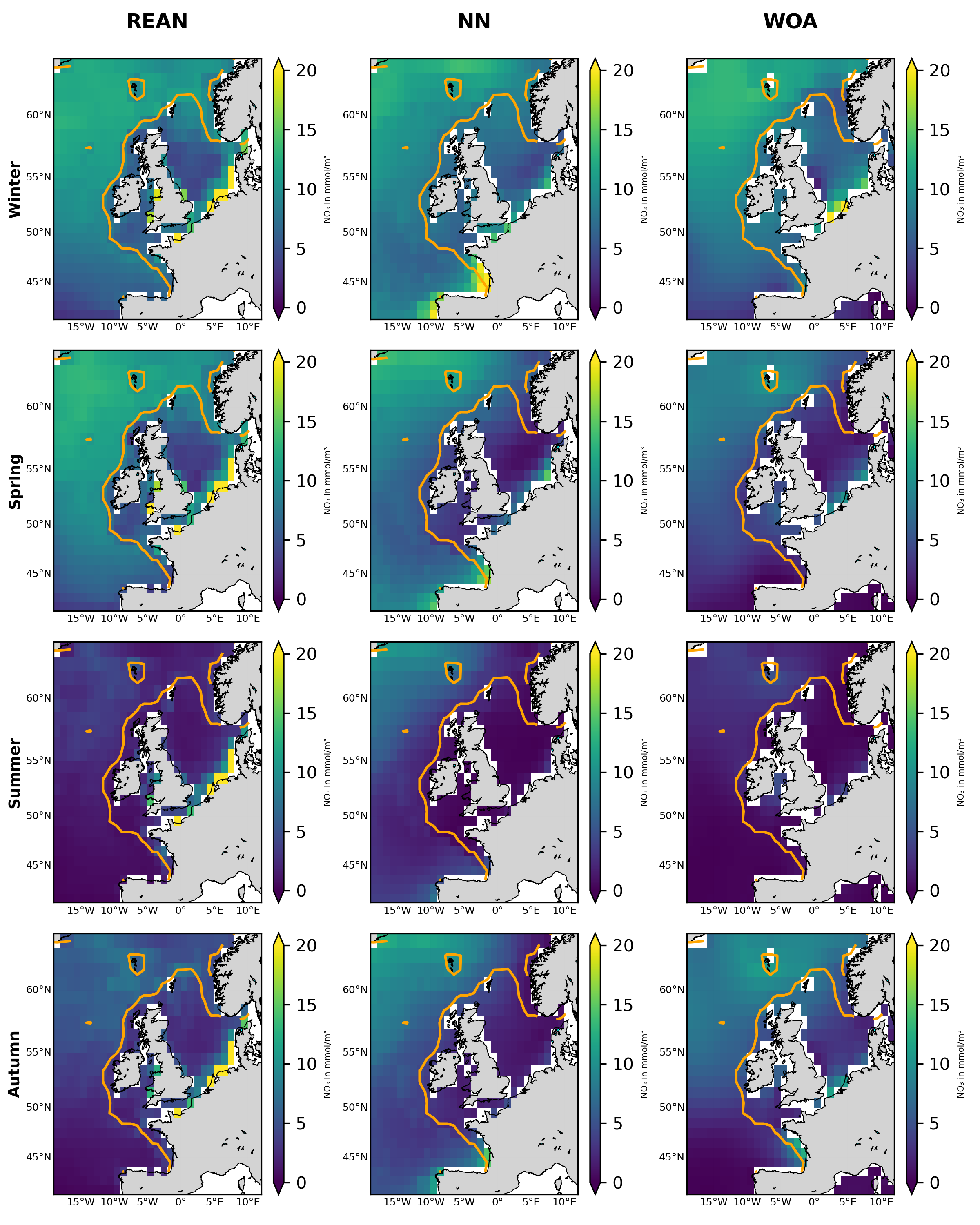}
\caption{Comparison of seasonal 1998-2020 surface nitrate climatology (in mmol/m$^{3}$) between the Copernicus reanalysis (left-hand column), the NN-predicted data-set assimilated in this study (middle column) and the World Ocean Atlas (WOA) data-set (right-hand column). It should be however noted that the WOA data-set is constructed from data taken from a much longer period than 1998-2020, i.e. starting in the early 20-th century \citep{garcia2019world}. The orange line shows the boundary of the Continental Shelf (bathymetry $<$ 200m).}
\label{Fig_1}
\end{figure}

A feed-forward neural network model (NN) was trained in \cite{banerjee2025improved} to predict NWES surface nitrate concentrations from a range of structural (e.g. coordinates, bathymetry), atmospheric (e.g. short-wave radiation, wind stress), riverine discharge inputs and variables from the ocean reanalysis with a very close proximity to satellite observations (i.e. SST, surface salinity, surface PFT chlorophyll, total surface net primary production and total surface phytoplankton carbon). For the NN inputs the training data came from the Copernicus Marine Service NWES reanalysis product NWSHELF\_\-MULTIYEAR\_\-BIO\_\-004\_\-011 \citep{kay2019north}, ERA5 atmospheric data, and the riverine discharge data from \cite{lenhart2010predicting}. The training data for nitrate came from the International Council for the Exploration of the Sea (ICES). (For more details see \citep{banerjee2025improved}.)  The NN model has been successfully used to produce a bi-decadal (1998-2020) gap-free, gridded daily and 7 km resolution dataset for surface nitrate across the NWES reanalysis domain. The NN model and the data-set showed good skill against independent observations \citep{banerjee2025improved}, however due to the relative simplicity of the NN model, the effective spatial and temporal resolution of the data-set has been shown to be coarser
than the data grid (about 30km spatial and 10 day temporal resolution). Furthermore, the nitrate data-set seasonal climatology compares well with the World Ocean Atlas (WOA, \cite{garcia2019world}), e.g. see Fig.5 of \citep{banerjee2025improved} and also Fig.\ref{Fig_1}. The NN-predicted nitrate also reveals the seasonal nitrate biases of the Copernicus reanalysis (see Fig.3 of \citep{banerjee2025improved} and Fig.\ref{Fig_1}), e.g. the persistent strong ($>$10mmol/m$^{3}$) positive bias in the southern North Sea, and a clear positive bias across nearly all of the NWES domain in the late Spring - early Summer (see also values at specific observing stations throughout NWES in \cite{banerjee2025improved}, or Fig.11 in \cite{kay2019north}).

In this work, we have assimilated the NN-generated surface nitrate values of \cite{banerjee2025improved} into the NEMO-FABM-ERSEM model. For methodological simplicity, we have assimilated daily nitrate values, to match the assimilation cycle of the other assimilated variables. To handle the coarser effective spatial resolution, as well as the potential impact of observational error spatial correlations, we have thinned the assimilated nitrate data to a 35 km spatial resolution scale. Since no observational error information was available, in this initial proof-of-concept work we used spatially and temporally constant background\-observation error ratio of 3:1, based on the average ratio found for chlorophyll in \citep{skakala2018assimilation}. Although this approach is considered sufficient at this initial prototype stage, it will need to be improved upon before any operational implementation. The nitrate assimilation updated only the modelled nitrate values, i.e. it was applied independently of the chlorophyll assimilation and associated balancing scheme.

\subsection{The experiments}


The system setup run in the experiments combined elements of both the Met Office operational forecasting system and the Copernicus, Met Office-produced, reanalysis of \cite{kay2019north}, with the main differences between the forecasting system, reanalysis and the runs from this study listed in Table \ref{tab:set-ups}. Due to the offline nature of this work and some existing differences in the Met Office reanalysis and forecasting systems, it is challenging to demonstrate the impact of nitrate assimilation in the Met Office operational forecasting suite, whilst ensuring complete consistency with the Copernicus reanalysis used to predict the assimilated nitrate. As described in Table \ref{tab:set-ups}, we have used a setup similar to the one used for operational forecasts, with certain tweaks to bring it closer to the Copernicus reanalysis set-up, such as introducing PFT chlorophyll-$a$ assimilation and assimilating the same version of satellite data as in the reanalysis. The hope is that this modelling choice would ensure that the experiments are ideally within reasonable proximity of both the nitrate-predicting reanalysis and the operational application used for forecasting. The drawback of this approach is that the analysis state in ML-nit DA deviates to a degree from the Copernicus reanalysis. As already mentioned, one of the goals of this study is to demonstrate the potential skill of a future ``online'' system, where nitrate is NN-predicted using inputs from the analysis state of the same run where it is being assimilated. If there is a major discrepancy between the analysis state in ML-nit DA and the reanalysis used by the NN-model, we risk that our ``offline'' system will significantly underestimate the skill of the future ``online'' system, as it lacks the full consistency of the online implementation. However, if such imperfect offline system substantially improves phytoplankton forecast skill through nitrate assimilation, it indicates that the impact of nitrate assimilation on phytoplankton forecasts is indeed robust. We have calculated the differences between the ML-nit DA analysis and the Copernicus reanalysis and estimated the size of the impact of those differences on our results. This is discussed in the Results section.

\begin{table}[h!]
\centering
\small
\begin{tabularx}{\textwidth}{lXXX}
\toprule
Set-up & \textbf{Operational} & \textbf{Reanalysis} & \textbf{This study} \\
\midrule
Assimilated SST & GHRSST & ESA CCI v1.1 & ESA CCI v1.1 \\

Assimilated in situ T \& S & GTS & ICOADS, EN4 & ICOADS, EN4 \\

Assimilated SLA & CMEMS & -- & -- \\

Assimilated chlorophyll-$a$ & total from CMEMS & PFT from ESA CCI v3.1 & PFT from ESA CCI v3.1 \\

Atmospheric forcing & Met Office NWP & ERA5 & Met Office NWP \\

Boundary conditions & Met Office NA model (in 2018) & GloSea reanalysis & Met Office NA model \\
\bottomrule
\end{tabularx}
\vspace{.25cm}
\caption{The main differences between the Met Office operational forecasting system, the 1997-2020 reanalysis of \cite{kay2019north, kay2021north} and the set-up from this paper.}
\label{tab:set-ups}
\end{table}


The three experiments in this study (no-nit DA, ML-nit DA, clim-nit DA) were performed for the biologically active period between March and September 2018, being initialized on the 01/03/2018 from the Copernicus reanalysis. The assimilation cycle in the experiments was daily, and at each day the model produced a separate 5-day forecast. As mentioned in the introduction, the ML-nit DA experiment assimilates the same data as no-nit DA plus the NN-derived surface nitrate from \cite{banerjee2025improved}. The clim-nit DA experiment replaces assimilation of the flow-dependent nitrate with assimilation of weekly varying surface nitrate climatology derived from the same 1998-2020 data of \cite{banerjee2025improved}.


\subsection{Skill metrics}

We have used a range of skill metrics to assess the assimilation as well as model forecast performance. Two of the metrics, ``the bias'' and ``the bias-corrected Root-Mean Square Error'' (RMSE) were defined:
\begin{equation}\label{bias}
\hbox{bias} = \langle \hbox{Model} \rangle - \langle \hbox{Observations} \rangle,
\end{equation}
and
\begin{equation}\label{BC-RMSE}
\hbox{bias-corrected RMSE} = \sqrt{\langle (\hbox{Model} - \hbox{Observations} - \hbox{bias})^{2} \rangle}.
\end{equation}
In the above the $\langle\rangle$ denote averaging (in this work typically through time).
Another metric used is RMSE skill improvement $\hbox{RMSE}_{imp}$, which is simply defined as 
\begin{equation}\label{RMSE-imp}
\hbox{RMSE}_{imp} =\hbox{RMSE}_{new} - \hbox{RMSE}_{ref}.
\end{equation}
$\hbox{RMSE}_{new}$ in the above means RMSE skill of a new product as measured relative to RMSE skill of a reference product ($\hbox{RMSE}_{ref}$). (The RMSE skill of both products is typically measured against observations, e.g. in this study against the assimilated satellite and NN-predicted data.)
We define also RMSE relative improvement, $\hbox{RMSE}_{rel-imp}$, as
\begin{equation}\label{RMSE-rel-imp}
\hbox{RMSE}_{rel-imp} = \frac{\hbox{RMSE}_{imp}}{\hbox{RMSE}_{ref}}.
\end{equation}
The $\hbox{RMSE}_{rel-imp}$ values vary between -1 and infinity, with negative values meaning RMSE improvement relative to the reference and positive values meaning RMSE degradation relative to the reference.

\subsection{Validation {\bf data} at the L4 station}

\begin{figure}
\hspace{0cm}
\noindent\includegraphics[width=13cm]{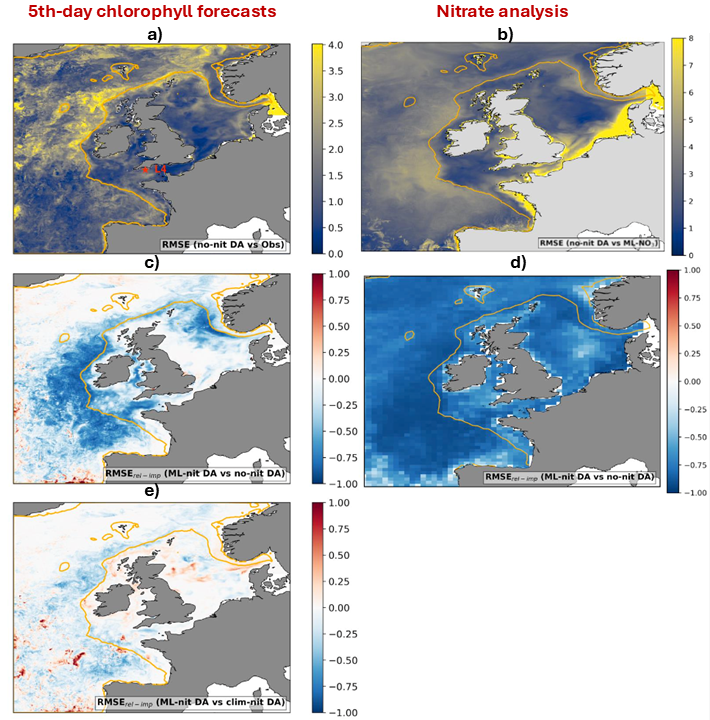}
\caption{The upper left-hand panel (a) shows the model 5-th day forecast skill in surface total chlorophyll-$a$ concentration when compared to the assimilated satellite OC product. The skill is measured by RMSE (in mg/m$^{3}$) calculated for each location through the simulation period. The upper right-hand panel (b) shows the same for analysis nitrate (comparing it to the assimilated ML-derived nitrate). The middle left-hand panel
(c) shows the improvement to the 5-th day forecast total chlorophyll-$a$ due to nitrate assimilation, as measured by the $\hbox{RMSE}_{rel-imp}$ metric (Eq.\ref{RMSE-rel-imp}), comparing ML-nit DA to no-nit DA (the negative values mean ML-nit DA outperforms no-nit DA). The middle right-hand panel (d) shows the same as (c), but for the analysis nitrate. The bottom left-hand panel (e) shows the same as (c), but comparing ML-nit DA and clim-nit DA, rather than ML-nit DA and no-nit DA. In the panel (a) we marked by the red-colored star the location of the L4 station used for in situ validation of the experiments from this study.}
\label{Fig_4}
\end{figure}

L4 station is operated by the Western Channel Observatory (WCO, https://\-www.\-westernchannelobservatory.\-org.uk/) in the western English Channel (50.25$^{\circ}$N, 4.217$^{\circ}$W) within the broader coastal zone 13km from the Plymouth Sound (see its location marked in Fig.\ref{Fig_4}:a). The location is relatively shallow (50m), within a region that is seasonally stratified and highly biologically productive \citep{pingree1978tidal}. The L4 station provides one of the longest time-series for a range of biogeochemistry variables worldwide, starting in 1988 \citep{harris2010l4}. This typically includes measurements for total chlorophyll-$a$ derived from fluorescence, data for nutrients (nitrate, phosphate, silicate, ammonium) and oxygen. The L4 observations are most abundant at/near the sea surface, but provided also for a range of depths across the water column. The different experiments from this study were validated using L4 data for chlorophyll-$a$, oxygen, nitrate, phosphate, silicate and ammonium. 

\section{Results and discussion}

\begin{figure}
\hspace{0.5cm}
\noindent\includegraphics[width=12cm]{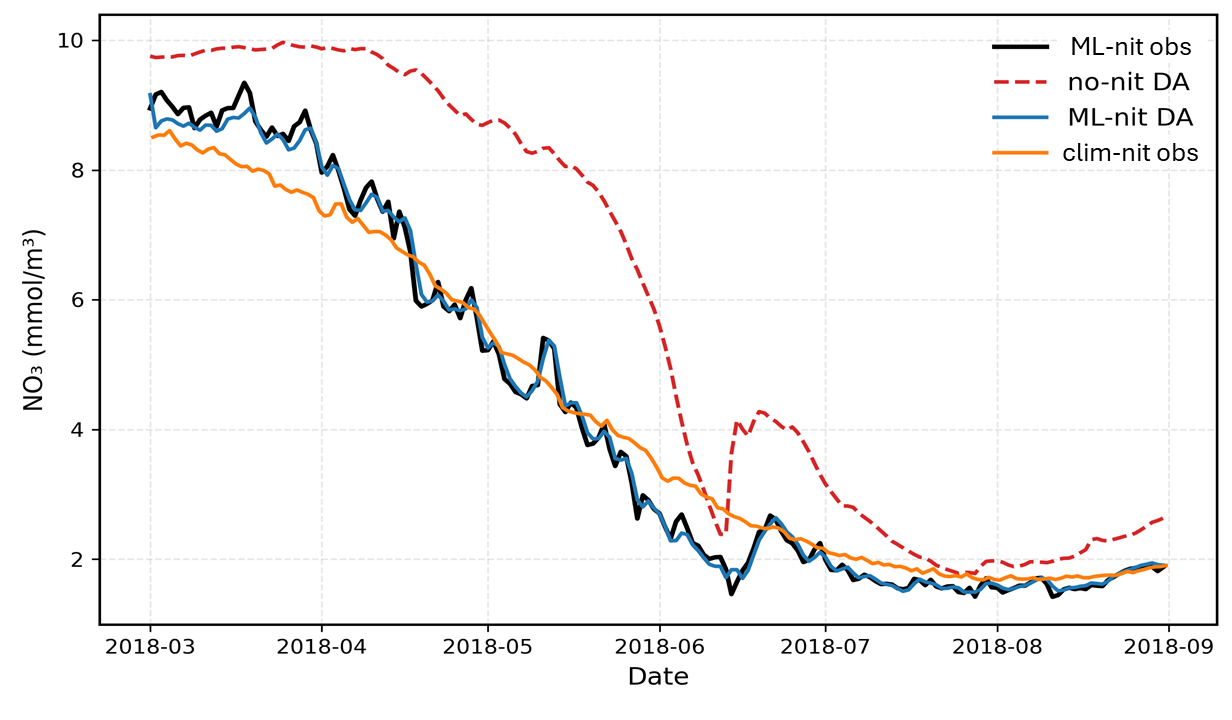}
\caption{Impact of NN-derived nitrate assimilation on the model surface nitrate as compared to the assimilated data. The domain-averaged surface nitrate time-series are being plotted for the NN-predicted nitrate (labelled ``ML-nit obs''), its weekly climatology (labelled ``clim-nit obs''), the no-nit DA run and the ML-nit DA run.}

\label{Fig_2}
\end{figure}

\begin{figure}
\hspace{-1cm}
\noindent\includegraphics[width=15cm]{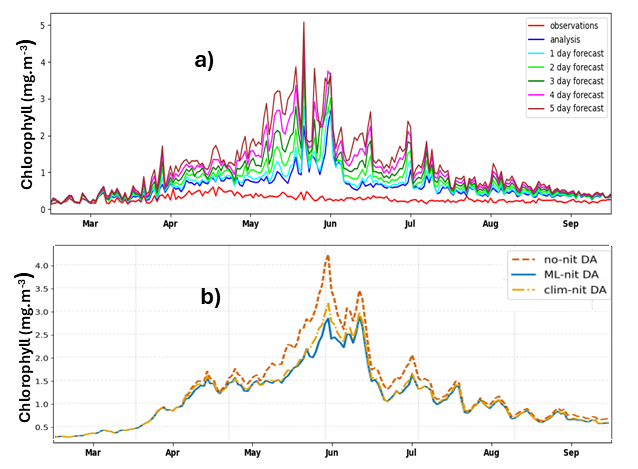}
\caption{ (a) The total surface chlorophyll-$a$ concentrations (in mg/m$^{3}$) averaged through the model domain for the analysis and the full range of forecasting days (1-5 day lead times). The no-nit DA model run is compared with the assimilated satellite OC-CCI observations (the OC satellite total chlorophyll-$a$ shown is a sum of the assimilated PFT chlorophyll-$a$ concentrations). The model data were masked wherever the observations had gaps in their values. (b) The domain-averaged time-series comparing the 5-th day forecasts across the three experiments from this study. Unlike the panel (a), here the averaging was not limited to observation locations, to show the full scale of the impact of nitrate assimilation.}
\label{Fig_3}
\end{figure}


As already discussed in Sec.2.2, unlike the NN-predicted surface nitrate, the reanalysis has significant NWES biases (mostly in late Spring - Summer) when compared to the WOA (Fig.\ref{Fig_1}). These biases are a long-term feature of the Met Office system, and it is therefore understandable that the same excess surface nitrate occurred also in the 2018 no-nit DA run from this study (Fig.\ref{Fig_2}). Fig.\ref{Fig_2} demonstrates that the largest overestimate of nitrate happens in the bloom-to-post bloom period in May-June. Fig.\ref{Fig_3}:a shows how these excess nitrate concentrations map into biases in phytoplankton chlorophyll-$a$ forecasts, i.e. it demonstrates the growth of the phytoplankton bias with the forecast lead-time in the no-nit DA run, with the largest skill degradation in the same May-June. The nitrate skill of the no-nit DA run is also shown through the RMSE metric in Fig.\ref{Fig_4}:b. It is clear from Fig.\ref{Fig_4}:a,b that those open sea regions which have the lowest nitrate skill (e.g. near the shelf-break) correspond well to the areas where the phytoplankton chlorophyll-$a$ 5-th forecasting day RMSE is the highest (Fig.\ref{Fig_4}:a). This indicates that the misrepresentation of nitrate is a key driver behind many biases in forecast chlorophyll. Curiously, this seems not to be the case in the southernmost part of the North Sea, which is an area of very large nitrate bias in the analysis (Fig.\ref{Fig_4}:b). The specific reasons why the southern North Sea is different are not clear and future effort will be dedicated to investigate this issue.

\begin{figure}
\hspace{0.5cm}
\noindent\includegraphics[width=12cm]{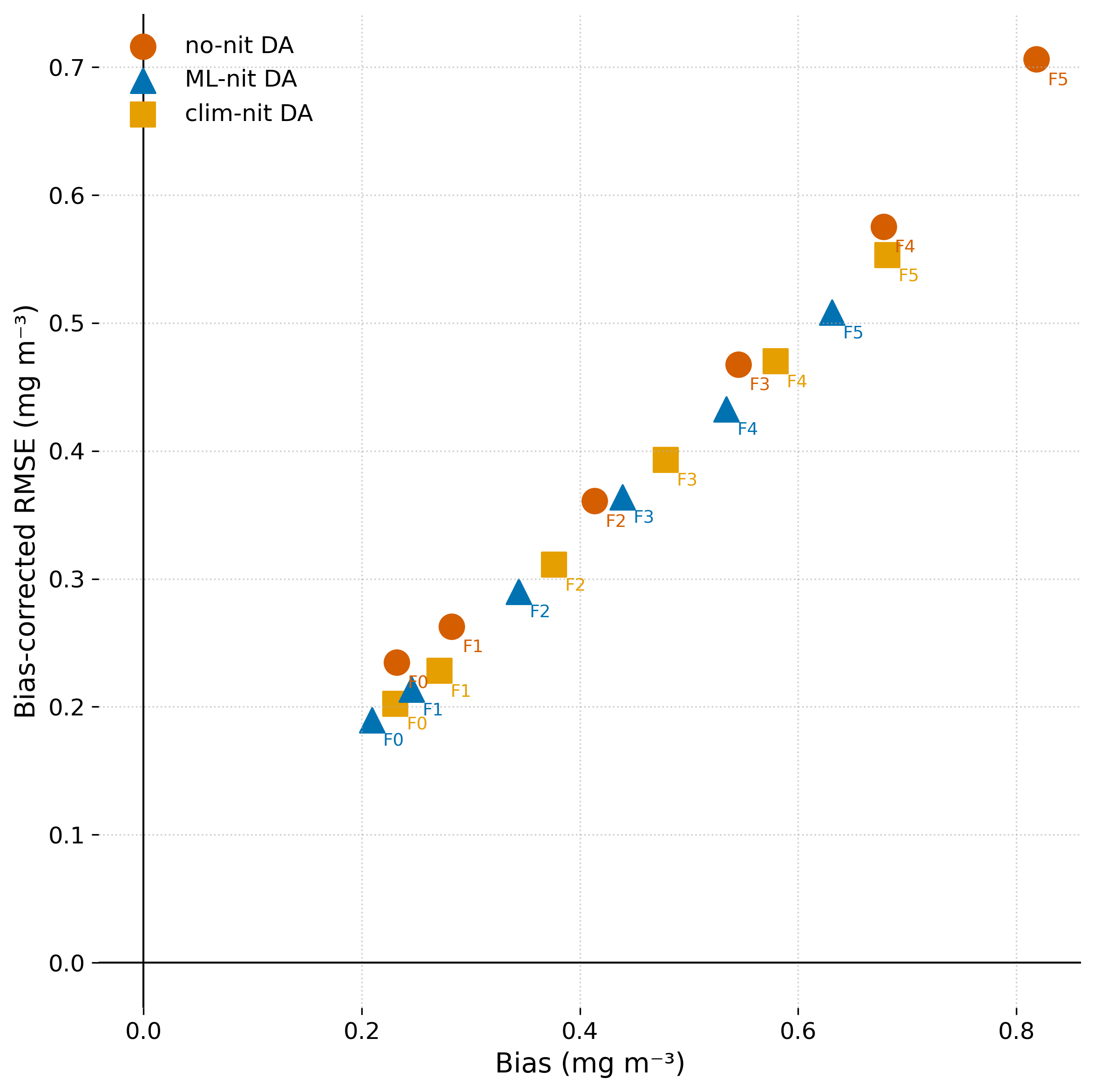}
\caption{Forecasting skill in surface total chlorophyll-$a$ concentrations relative to the assimilated satellite OC observations (as before, the satellite OC total chlorophyll-$a$ being taken as the sum of the assimilated PFT chlorophyll-$a$). The x-axis shows the bias as defined in Eq.\ref{bias} and the y-axis BC-RMSE as defined in Eq.\ref{BC-RMSE} (for more detail see Sec.2.4). The different forecast lead times (days) are marked as ``F0-5'', with ``0'' standing for analysis.}
\label{Fig_5}
\end{figure}


Fig.\ref{Fig_4}:d also demonstrates that assimilating nitrate into the model achieves its stated purpose, i.e. it substantially reduces nitrate RMSE relative to the assimilated data-set (reducing RMSE shown in Fig.\ref{Fig_4}:b). This has then desirable impact on the 5-th day phytoplankton chlorophyll-$a$ forecast (RMSE$_{rel-imp}$ in Fig.\ref{Fig_4}:c), improving the 5-th day forecast largely in the regions where the nitrate correction is most significant (RMSE$_{rel-imp}$ in Fig.\ref{Fig_4}:d). In terms of the nitrate seasonal biases, nitrate assimilation is shown to remove the excess nitrate in the Spring-Summer period (see Fig.\ref{Fig_2}). In the same period, during the peak and the recession of the bloom, we observe the largest impact of nitrate assimilation on the phytoplankton forecast bias (Fig.\ref{Fig_3}:b). A detailed insight into the model skill across the full forecasting period is shown in Fig.\ref{Fig_5}. Fig.\ref{Fig_5} demonstrates that both the bias and BC-RMSE are improved consistently across the whole 5-day forecasting period, with the improvement increasing with the forecast lead time.


A key question that needs exploring is how much benefit there is from assimilating nitrate time-evolving values, as opposed to relying on nitrate (ML-derived) seasonal climatology (for their difference see Fig.\ref{Fig_2}), which can always be supplied ``offline'' with lower computational cost. Fig.\ref{Fig_5} shows that clim-nit DA significantly improves the forecast compared to the no-nit DA run, but not as much as the ML-nit DA run, with the difference in their performance steadily growing with lead time. The ML-nit DA run performs better than the clim-nit DA run on large parts of the outer NWES boundary with RMSE improvement (RMSE$_{imp}$) broadly in the range of 10-50\% (Fig.\ref{Fig_4}:e). The relative RMSE degradation with ML-nit DA relative to clim-nit DA happens on much smaller areas of the domain than the improvement, even though in some very specific locations the degradation can be quite substantial (as high as 100\%, see Fig.\ref{Fig_4}:e).

Furthermore, to evaluate the limitations of the offline system implemented within ML-nit DA (the nitrate has been predicted from the Copernicus reanalysis rather than from the ML-nit DA analysis state), we have calculated the differences between the (assimilated) nitrate predicted by NN from the Copernicus reanalysis and the same nitrate predicted by NN from the ML-nit DA run. Our analysis (not shown here) demonstrated that the predicted nitrate difference (measured by RMSD) was around 30\% smaller than the difference between the assimilated nitrate and its weekly climatology. Based on this we conclude that the difference between the phytoplankton forecast skill of the online and the offline systems would be smaller than the difference between clim-nit DA and ML-nit DA shown in Fig.\ref{Fig_5}. We would also conjecture that assimilating nitrate in the online system might further improve the phytoplankton forecast relative to ML-nit DA, as it is more self-consistent than the offline assimilation. These conjectures however, need to be demonstrated to hold true when such a system is developed in the future.

\begin{figure}
\hspace{-.5cm}
\noindent\includegraphics[width=15cm]{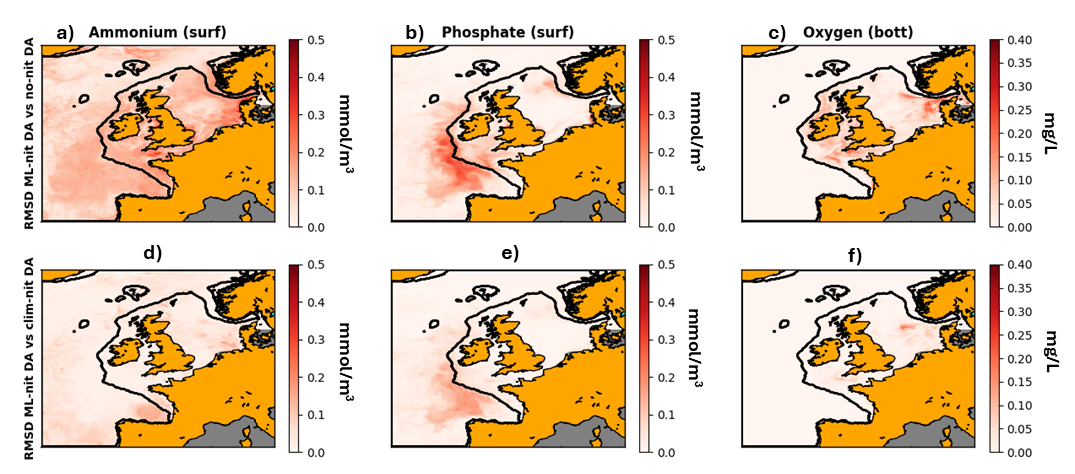}
\caption{Impact of nitrate assimilation on a range of variables included in the OSPAR eutrophication indicators: sea surface ammonium (left-hand column, a), d), mmol/m$^{3}$) and phosphate (middle column, b), e), mmol/m$^{3}$), and sea bottom oxygen (right-hand column, c), f), mg/L). The impact is measured through Root Mean Square Difference (RMSD) between the ML-nit DA run and no-nit DA run (upper row) and the ML-nit DA run and the clim-nit DA run (bottom row). The RMSD is calculated for each location across the simulation period. The comparison is done only for the 5-th forecast day.}
\label{Fig_6}
\end{figure}

\begin{figure}
\hspace{-.5cm}
\noindent\includegraphics[width=14cm, height=13cm]{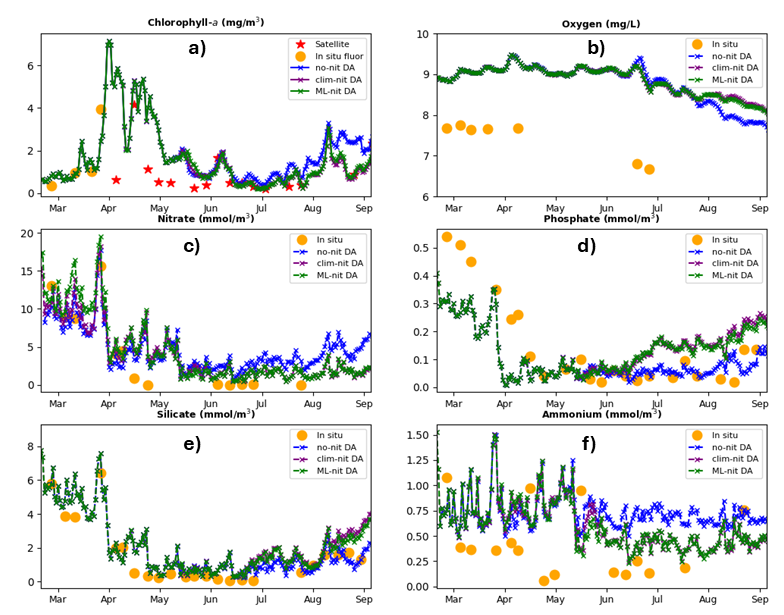}
\caption{The variables important for eutrophication-relevant indicators compared with observations at the L4 station. The comparison is done only for the 5-th day forecast. The variables compared are (a) total surface chlorophyll-$a$, (b) sea bottom oxygen, (c) surface nitrate, (d) surface phosphate, (e) surface silicate and (f) surface ammonium.}
\label{Fig_7}
\end{figure}

One potentially important application of short-range NWES forecasts is predicting eutrophication events. The improvement in model phytoplankton forecast skill, along with improved forecast nitrate concentrations, should significantly contribute to the operational capability to forecast such events. Furthermore, when it comes to capturing extreme events, such as eutrophication, there is an obvious advantage in predicting time-evolution of nitrate ``online'' by ML, as opposed to using nitrate climatology. In Fig.\ref{Fig_6} we focus on some variables related to a range of standard eutrophication indicators beyond chlorophyll-$a$ (e.g. see OSPAR report \cite{axe2017eutrophication}): (i) dissolved surface inorganic nitrogen, represented in the model by the sum of nitrate and ammonium, (ii) dissolved surface inorganic phosphorus, represented in the model by phosphate, and (iii) dissolved oxygen near the sea bottom. Fig.\ref{Fig_6} shows the difference nitrate assimilation makes to 5-th forecasting lead day prediction of some key variables related to these indicators. It can be seen that the difference is quite significant especially for ammonium, which is understandable, since nitrogen cycling has been significantly altered through the assimilation of nitrate. The difference to phosphate is mainly outside of coastal zones, so it has lower impact on eutrophication monitoring, whereas the difference to the dissolved oxygen at the sea bottom is overall relatively small, but occurs mainly in certain interesting coastal areas, including the south-east North Sea (near the Danish coastline), which has seen hypoxia previously \citep{topcu2015seasonal}. Fig.\ref{Fig_6} also shows that the difference (measured by RMSD) between ML-nit DA and clim-nit DA is significantly smaller (roughly 5 times) than between ML-nit DA and no-nit DA.

Fig.\ref{Fig_7} validates the skill of the 5-th day model forecast of some key eutrophication indicators at the L4 observing station in the western English Channel. Unfortunately, although the overall impact of nitrate assimilation on phytoplankton chlorophyll-$a$ forecasting was large in the western English Channel (Fig.\ref{Fig_4}:c), this area of large impact excludes the coastal area where L4 is located. As shown in Fig.\ref{Fig_7}, there is a distinctive (generally positive) impact of nitrate assimilation on the nitrate in March-April and the Summer period. Consistent with Fig.\ref{Fig_3}:b there is little impact of nitrate assimilation on chlorophyll-$a$ forecast in March-April, but there is more significant and positive impact in the Summer. Unlike the domain-wide results where the nitrate assimilation impact on chlorophyll-$a$ forecast is mostly visible around June (Fig.\ref{Fig_3}:b), here it becomes larger as the simulation progresses. The progressive shift in chlorophyll-$a$ forecast (Fig.\ref{Fig_7}) triggers changes in the other nutrients (phosphate, ammonium and silicate), which in some cases improve forecast skill (ammonium) and in others degrade it (phosphate and silicate). The improvement in ammonium forecast is however particularly interesting as it is part of a broader improvement in forecasting inorganic nitrogen (in the model represented by nitrate and ammonium).

\section{Conclusions}

In this work, we have demonstrated that a combined (hybrid) machine learning - data assimilation (DA) system where surface nitrate is being predicted by a neural network (NN) from the model analysis state (as well as atmospheric, structural and riverine data) and subsequently assimilated into the model, can have a major positive impact on phytoplankton short-range (up to 5 day) forecasts in a shelf sea environment. We have argued that this happens because the degradation to phytoplankton forecast skill is due to an imbalance between the simulated light and nutrients, triggered by the lack of update to nutrients in the assimilation step within the existing operational system. We have shown that although a significant improvement to the phytoplankton forecast skill can be achieved through assimilating the NN-derived surface nitrate weekly climatology, the flow-dependent prediction of nitrate outperforms the climatology approach. We have also evaluated the broader impact of nitrate assimilation on the forecast of a wider range of eutrophication indicators and performed some validation of this impact at the L4 location. 

This work is complementary to other current attempts to combine ML with DA to improve the Met Office operational system and short-term forecasts, e.g. \citep{higgs2025hybrid}. In the work of \cite{higgs2025hybrid} ML was used to learn the cross-covariances in the background error covariance matrix as an alternative to expensive ensemble methods and also to predict directly increments of unobserved variables from the increments of chlorophyll in an end-to-end approach. Unlike \cite{higgs2025hybrid} we leave here the DA scheme unchanged, but instead let ML supply the observations for the assimilation. Similar to \cite{higgs2025hybrid}, we anticipate that the technique developed here might have important use in future operational forecasting delivered by the Met Office for the North-West European Shelf. For this the NN-based nitrate assimilation will need to be implemented online, and other aspects of the DA system should be ideally improved upon, such as the estimates of the background and observation nitrate errors. Ensembles of ML models themselves can help with the uncertainty estimates of the assimilated nitrate, being computationally cheap ways to address epistemic uncertainty, but less so aleatory uncertainty (e.g. \cite{hullermeier2021}). It is therefore likely that any such ML-based uncertainty estimates would need improving upon, e.g. by using diagnostic methods such as of \citep{desroziers2005diagnosis}.

In future work we will also look to expand our present approach to include other important variables currently not updated by the assimilation system, such as phosphate and oxygen. Another update that we envision for the future is to improve the spatial and temporal resolution of the NN-predicted nitrate by increasing the complexity of the NN model (as discussed in \cite{banerjee2025improved}). This could bring additional benefit for the phytoplankton forecast and also increase the relative benefit of flow-dependent prediction of nitrate compared to assimilating nitrate climatology. Finally, we plan to test using the gap-free ML-prediction of nitrate more directly in a bias-correction scheme, rather than within the assimilation framework.

The methods presented here demonstrate that implementing machine learning within DA offers a cheaper and skilled alternative to using expensive ensemble techniques such as ensemble Kalman filters to provide multivariate updates from assimilation of observed variables.

\vspace{.5cm}

{\bf Acknowledgments}  
We acknowledge use of the Monsoon2 system, a collaborative facility supplied under the Joint Weather and Climate Research Programme, a strategic partnership between the Met Office and the UK Natural Environment Research Council (NERC). The model runs used river data prepared by Sonja van Leeuwen and Helen Powley as part of UK Shelf
Sea Biogeochemistry programme (contract no.NE/K001876/1) of the NERC and the Department for Environment, Food and Rural Affairs (DEFRA). The riverine data contained also climatological values from the Global River Discharge Data Base and the Centre for Ecology and Hydrology \citep{young2007prediction}. 

{\bf Funding:}  This work was partly funded by the Horizon Europe project The New Copernicus Capability
for Tropic Ocean Networks (NECCTON, grant agreement no.101081273). We also acknowledge support from the UK NERC, including the single centre national capability programme – Climate Linked Atlantic Sector Science (CLASS,379NE/R015953/1).

{\bf Conflict of Interest statement:} The authors declare no conflict of interest.

{\bf Permission to reproduce material from other sources:} No material from other sources was reproduced in this work.

{\bf Authors’ contributions:} DB contributed Software, Investigation, Visualisation, Formal Analysis, Writing – review and editing. JS contributed 
Conceptualization, Supervision, Formal Analysis, Visualisation, Writing – Original draft, Writing – review and editing. DF contributed Software, Investigation, Writing – review and editing. 

{\bf Data availability statement:} The ML model can be downloaded from https://\-github.com/\-neccton-algo/\-NO3\_Emulator\_NECCTON. The simulation outputs are stored on the MonSOON facility MASS and can be obtained upon request. The assimilated nitrate product can be downloaded from https://zenodo.org/records/19695959 and visualised in detail through https://ml-nitrate-viewer-banerjee.streamlit.app/. 

\end{document}